\documentclass[twocolumn,times]{aastex6}
\usepackage{amsmath}
\usepackage{multirow}
\usepackage[nolist]{acronym}

\begin{document}

\title{Going the Distance: Mapping Host Galaxies of LIGO and Virgo Sources in Three Dimensions Using Local Cosmography and Targeted Follow-up}
\shorttitle{Advanced LIGO/Virgo Volume Reconstruction and Galaxy Catalogs}

\slugcomment{The Astrophysical Journal Letters, 829:L15, 2016 September 20}
\received{2016 March 23}
\revised{2016 July 28}
\accepted{2016 July 29}
\published{2016 September 21}

\AuthorCallLimit=-1

\author{Leo~P.~Singer\altaffilmark{1,2}}
\author{Hsin-Yu~Chen\altaffilmark{3}} \author{Daniel~E.~Holz\altaffilmark{3}} \author{Will~M.~Farr\altaffilmark{4}} \author{Larry~R.~Price\altaffilmark{5}} \author{Vivien~Raymond\altaffilmark{5,6}} \author{S.~Bradley~Cenko\altaffilmark{1,7}} \author{Neil~Gehrels\altaffilmark{1}} \author{John~Cannizzo\altaffilmark{1}} \author{Mansi~M.~Kasliwal\altaffilmark{8}} \author{Samaya~Nissanke\altaffilmark{9}} \author{Michael~Coughlin\altaffilmark{10}} \author{Ben~Farr\altaffilmark{3}} \author{Alex~L.~Urban\altaffilmark{11}} \author{Salvatore~Vitale\altaffilmark{12}} \author{John~Veitch\altaffilmark{4}} \author{Philip~Graff\altaffilmark{13}}
\author{Christopher~P.~L.~Berry\altaffilmark{4}} \author{Satya~Mohapatra\altaffilmark{12}} \author{Ilya~Mandel\altaffilmark{4}} 
\altaffiltext{1}{Astroparticle Physics Laboratory, NASA Goddard Space Flight Center, Mail Code 661, Greenbelt, MD 20771, USA}
\altaffiltext{2}{NASA Postdoctoral Program Fellow}
\altaffiltext{3}{Department of Physics, Enrico Fermi Institute, and Kavli Institute for Cosmological Physics, University of Chicago, Chicago, IL 60637, USA}
\altaffiltext{4}{School of Physics and Astronomy, University of Birmingham, Birmingham, B15 2TT, UK}
\altaffiltext{5}{LIGO Laboratory, California Institute of Technology, Pasadena, CA 91125, USA}
\altaffiltext{6}{Albert-Einstein-Institut, Max-Planck-Institut f\"ur Gravitationsphysik, D-14476 Potsdam-Golm, Germany}
\altaffiltext{7}{Joint Space-Science Institute, University of Maryland, College Park, MD 20742, USA}
\altaffiltext{8}{Cahill Center for Astrophysics, California Institute of Technology, Pasadena, CA 91125, USA}
\altaffiltext{9}{Institute of Mathematics, Astrophysics and Particle Physics, Radboud University, Heyendaalseweg 135, 6525 AJ Nijmegen, The Netherlands}
\altaffiltext{10}{Department of Physics and Astronomy, Harvard University, Cambridge, MA 02138, USA}
\altaffiltext{11}{Leonard E. Parker Center for Gravitation, Cosmology, and Astrophysics, University of Wisconsin--Milwaukee, Milwaukee, WI 53201, USA}
\altaffiltext{12}{LIGO Laboratory, Massachusetts Institute of Technology, 185 Albany Street, Cambridge, MA 02139, USA}
\altaffiltext{13}{Department of Physics, University of Maryland, College Park, MD 20742, USA}

\shortauthors{Singer et al.}


\newcommand{\dd}{\ensuremath{\mathrm{d}}}

\defcitealias{GoingTheDistance}{Letter}
\defcitealias{GoingTheDistanceSupplement}{Supplement}
 
\begin{abstract}
The Advanced \ac{LIGO} discovered \acp{GW} from a binary black hole merger in 2015 September and may soon observe signals from neutron star mergers. There is considerable interest in searching for their faint and rapidly fading \ac{EM} counterparts, though \ac{GW} position uncertainties are as coarse as hundreds of square degrees. Since LIGO's sensitivity to binary neutron stars is limited to the local Universe, the area on the sky that must be searched could be reduced by weighting positions by mass, luminosity, or star formation in nearby galaxies. Since \ac{GW} observations provide information about luminosity distance, combining the reconstructed volume with positions and redshifts of galaxies could reduce the area even more dramatically. A key missing ingredient has been a rapid \ac{GW} parameter estimation algorithm that reconstructs the full distribution of sky location and distance. We demonstrate the first such algorithm, which takes under a minute, fast enough to enable immediate electromagnetic follow\nobreakdashes-up. By combining the three\nobreakdashes-dimensional posterior with a galaxy catalog, we can reduce the number of galaxies that could conceivably host the event by a factor of 1.4, the total exposure time for the \emph{Swift} X\nobreakdashes-ray Telescope by a factor of 2, the total exposure time for a synoptic optical survey by a factor of 2, and the total exposure time for a narrow\nobreakdashes-field optical telescope by a factor of 3. This encourages us to suggest a new role for small field of view optical instruments in performing targeted searches of the most massive galaxies within the reconstructed volumes.
\end{abstract}

\acresetall
\acused{HEALPix}
\acused{2D}
\acused{3D}

\section{Introduction}
\label{sec:introduction}

The Advanced \acl{LIGO} (\acs{LIGO}; \citealt{AdvancedLIGO}) began operations in 2015 \citep{GW150914-DETECTORS} and almost immediately recorded the first-ever \ac{GW} signal from a \ac{BBH} merger, GW150914 \citep{GW150914-DETECTION}. It should soon observe \acp{GW} from \ac{NS} binary mergers \citep{LIGORates,O1-BNS-NSBH} too. These systems should present several kinds of \ac{EM} transients that are detectable by existing and planned facilities (e.g., \citealt{MostPromisingEMCounterpart}). Joint broadband observations would tell the full story of these rare events and solve longstanding puzzles from the nature of \aclp{SGRB} (\acsp{SGRB}\acused{SGRB}; \citealt{1986ApJ...308L..43P,1989Natur.340..126E,1992ApJ...395L..83N,2011ApJ...732L...6R}) to the astrophysical sites of $r$-process nucleosynthesis (\citealt{2014MNRAS.439..744R,2015MNRAS.447..140V}; etc.) and enable these systems to be used as standard siren probes of the evolution history of the universe \citep{StandardSirens1986,StandardSirens2006,StandardSirens2010}. \citet{MostPromisingEMCounterpart} consider the radioactively powered \acl{KN} \citep{kilonova} to be the most promising \ac{EM} signature to find in coincidence with a \ac{LIGO} event, though \citealt{BarnesKasenKilonovaOpacities} have shown that the high optical opacities of the ejecta will cause this signature to be faint ($M_i \lesssim -13$) and red ($r - i \sim 1$) and to peak quickly, within days to weeks (e.g. \citealt{KilonovaDiscWindsOutflows}). A consortium of partner gamma-ray, X-rays, optical, and radio facilities have embarked on an unprecedented campaign to search for \ac{EM} counterparts of \ac{GW} signals \citep{S6LowLatencyImplementationAndTesting,SwiftFollowupGWTransients,S6Optical,GW150914-EMFOLLOW}.

\ac{GW} localizations are currently $\sim 100$--$1000$\,deg$^2$ \citep{NissankeKasliwalEMCounterparts,FirstTwoYears,FirstTwoYearsRecolored,BurstFirstTwoYears} and should shrink to $\sim 10$\nobreakdashes--$100$\,deg$^2$ \citep{LIGOObservingScenarios} over years of detector upgrades and construction of additional detectors: Advanced Virgo \citep{aVirgo}, KAGRA, and LIGO\nobreakdashes--India. Even the most accurate imaginable \ac{GW} localizations of $\lesssim$10\,deg$^2$ will be grossly larger than the $\sim$1$\arcmin$--10$\arcmin$ \acp{FOV} of CCD cameras that are common on the world's largest optical and infrared telescopes. Consequently, robotic and low\nobreakdashes-overhead synoptic survey telescopes with primary mirror diameters of 1\nobreakdashes--8\,m and \acp{FOV} of up to tens of deg$^2$ have been heralded as the most promising tools \citep{MostPromisingEMCounterpart} for finding those fast and faint \ac{EM} counterparts.

\citet{CBCG} and \citet{GWGC} recognized that targeting individual galaxies can reduce the area to be searched while eliminating false positive candidates. A galaxy catalog was central to the follow\nobreakdashes-up strategy during Initial \ac{LIGO} \citep{CBCLowLatency,SwiftFollowupGWTransients,S6Optical}. Although the increasing number of galaxies within the expanding \ac{GW} range decreases the utility of this technique, \citet{UtilityGalaxyCatalogsWideFieldTelescopes} and \citet{KasliwalTwoDetectors} argued that using both sky positions and distance estimates from \acp{GW} can still reduce the number of galaxies, especially in the early years when localizations are particularly coarse. \citet{NissankeKasliwalEMCounterparts} showed that the region bracketed by the upper and lower limits of the 95\% credible distance interval can reduce the volume, and hence the number of galaxies, by a factor of 60\%.

While the Advanced \ac{LIGO} commissioning plan calls for several steps in sensitivity and range, parallel construction of additional detectors will result in shrinking sky localization uncertainty \citep{Veitch:2012,RodriguezBasicParameterEstimation}. Intriguingly, \citet{GalaxyStrategy} point out that these effects roughly cancel each other, such that the typical \ac{GW} error volume may scarcely vary over the next decade.

The obvious next step is to exploit the correlated \ac{3D} structure of the reconstructed volumes. This immediately raises three questions: How accurately can distance be measured with \ac{GW} detectors, particularly in the early two\nobreakdashes-detector configurations? What is the \ac{3D} shape of the reconstructed volumes? What is the minimal amount of information that is needed to faithfully describe these volumes?

In \citet{FirstTwoYears}, we elucidated the localization areas and shapes that we expect in early Advanced \ac{LIGO} and Virgo. In a similar spirit and using the same catalog of simulated events, we now reveal the shape, scale, and overall character of the \ac{3D} reconstructed volumes---enabled by a new and extremely efficient encoding of the \ac{3D} probability distributions. We provide a representative sample of \ac{GW} volume reconstructions in a format that could be made available beginning with the second Advanced \ac{LIGO} observing run. Complementing existing technologies to detect \citep{Cannon:2011vi,GW150914-GSTLAL} and localize \citep{BAYESTAR} \ac{GW} mergers within minutes of data acquisition, our approach can provide distance-resolved \ac{GW} sky maps in near real time.

All classes of instruments can benefit from the new \ac{3D} localizations, but they are especially powerful for conventional, large-aperture telescopes with narrow-\ac{FOV} instruments, and particularly at near infrared wavelengths, where large-\ac{FOV} cameras are scarce \citep{VISTA}. Our galaxy-targeted strategy of monitoring the most probable $\sim$100 galaxies for several nights following a \ac{GW} trigger could be implemented on large infrared facilities. However, even a pilot program on robotic 2\,m telescopes could be surprisingly powerful for the first few Advanced \ac{LIGO}\nobreakdashes--Virgo observing runs.

\begin{figure}
  \includegraphics[width=\columnwidth]{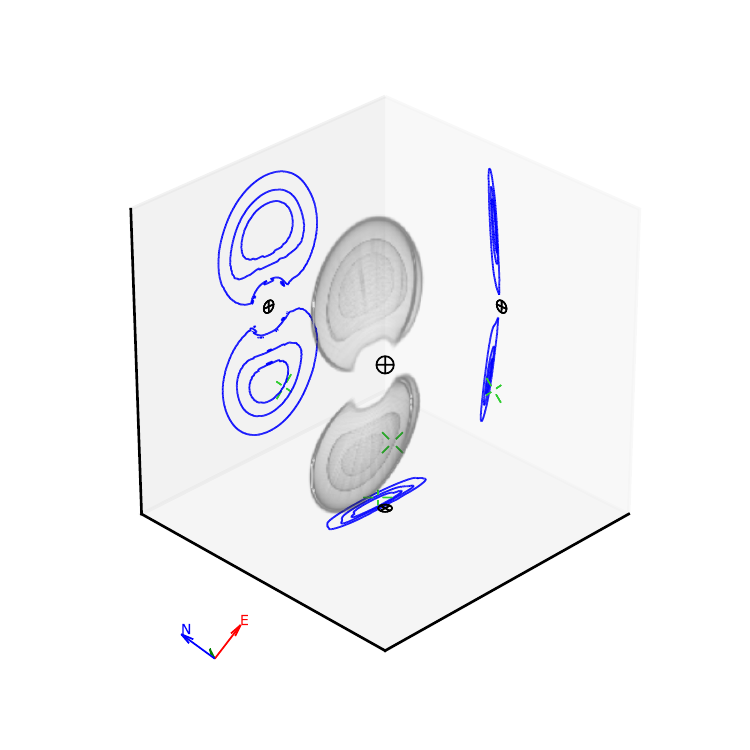}
  \caption{\label{fig:volume-rendering}Volume rendering of the 20\%, 50\%, and 90\% credible levels of a typical two-detector early Advanced \ac{LIGO} event. The three planes are perpendicular to the principal components of the probability distribution. The observer's position (the Earth) is at the origin. The green reticle shows the true position of the source. The compass in the bottom left corner shows the basis vectors of the equatorial coordinate system.}
\end{figure}

\begin{figure*}
    \includegraphics[width=\columnwidth]{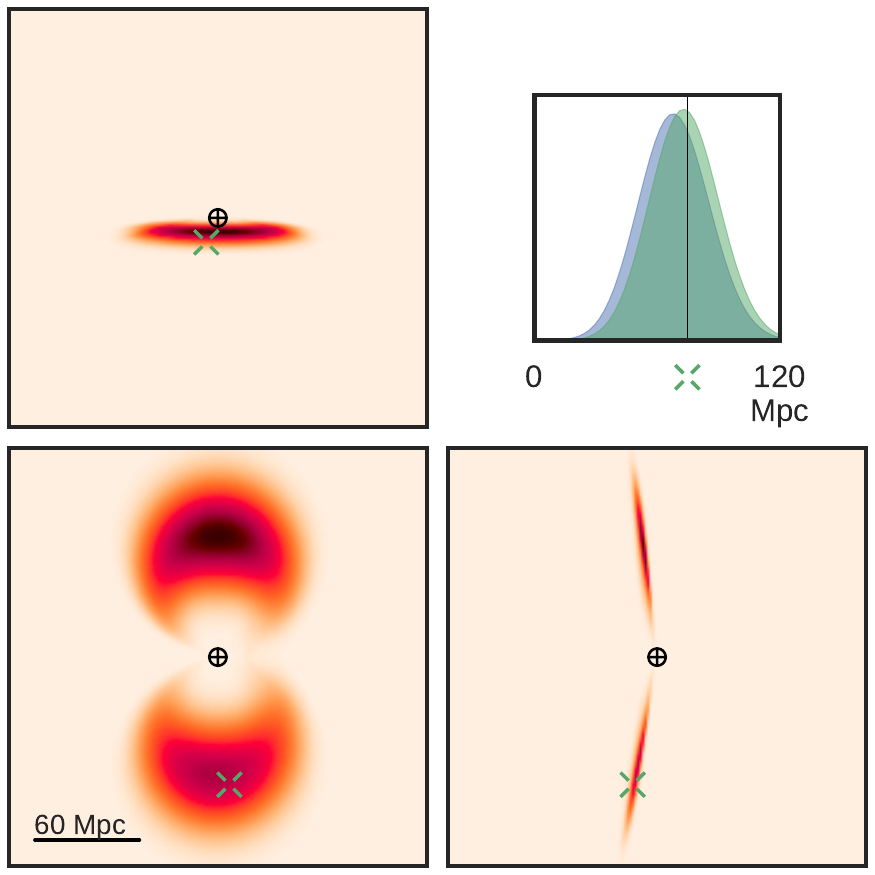}
    \hfill
    \includegraphics[width=\columnwidth]{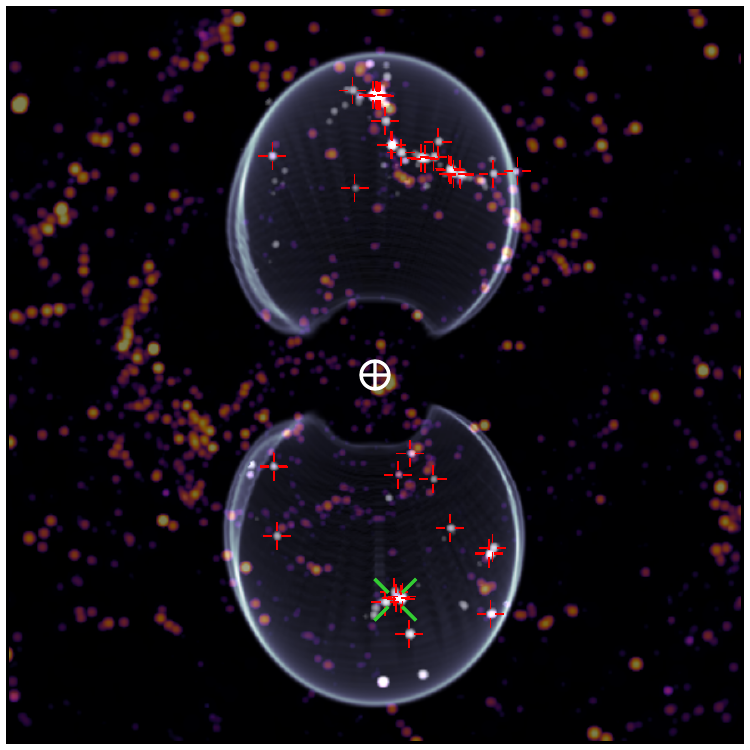}
    \caption{\label{fig:volume}Left panel: marginal posterior probability distribution in the principal planes, as in Fig.~\ref{fig:volume-rendering}. The inset shows the marginal distance posterior distribution integrated over the whole sky (blue) and the conditional distance posterior distribution in the true direction of the source (green). Right panel: volume rendering of the 90\% credible region superimposed over a slice of the galaxy group map of \citet{2MASSGalaxyGroups}. The most massive galaxies inside the credible region are highlighted.}
\end{figure*}

\section{Distance constraints}
\label{sec:distance-constraints}

The amplitude, or \ac{SNR}, of a \ac{GW} signal is determined by a degenerate combination of inclination and distance.\footnote{We do not distinguish between different cosmological distance measures; the direction\nobreakdashes-averaged \ac{BNS} range of Advanced \ac{LIGO} is $\sim 200$\,Mpc, or $z \sim 0.05$.} The Malmquist bias leads to a broad universal distribution of binary inclination angles, peaking at $30\arcdeg$ \citep{ShutzThreeFiguresOfMerit}. The distance of a \ac{GW} source can generally be estimated with $\sim 30\%$ fractional uncertainty \citep{1994PhRvD..49.2658C,FirstTwoYearsRecolored}.

The \emph{effective distance} $r_\mathrm{eff}$ of a \ac{GW} signal is the maximum distance at which it could have produced the observed \ac{SNR}. The \emph{horizon distance} $r_\mathrm{H}$ of a detector is the farthest distance at which the most favorably oriented source (at the detector's zenith, and with a binary inclination of $\iota = 0\arcdeg$) would register a threshold \ac{SNR} (generally defined as $\text{\ac{SNR}} = 8$). We give approximate formulae for the horizon distance in Equations~(1) and~(2) of \citet{FirstTwoYears}. The \emph{range} is the direction- and orientation\nobreakdashes-averaged distance of sources detectable at a threshold \ac{SNR}, $r_\mathrm{R} \approx r_\mathrm{H} / 2.26$ \citep{1993PhRvD..47.2198F,ShutzThreeFiguresOfMerit}.

The effective distance description has two major limitations. First, the source may sometimes lie beyond the effective distance because of measurement noise. Worse, there is no obvious way to describe the probability enclosed within, say, the \ac{2D} 90\% credible region on the sky and the effective distance; this number is always $\leq 90$\%. Second, notwithstanding the large fractional distance uncertainty, there is nontrivial structure to the full \ac{3D} reconstructed volumes that can be exploited to reduce the volume under consideration.

During \ac{O1}, the network \citep{LIGOObservingScenarios} consisting of \ac{LHO} and \ac{LLO} tends to produce probability sky maps consisting of one to two long, thin sections of a great circle \citep{KasliwalTwoDetectors,FirstTwoYears}. We provide this as an illustration of the main features for a two-detector network. We assume, as in \citep{LIGOObservingScenarios}, a \c{BNS} range of 54\,Mpc, though the range during \ac{O1} was about 40\% better. The corresponding \ac{3D} geometry is shown in Figs.~\ref{fig:volume-rendering}~and~\ref{fig:volume}. The degenerate arcs correspond to either one or two thin, rounded, slightly oblique petals, about 1$\arcdeg$--5$\arcdeg$ wide, 10$\arcdeg$--100$\arcdeg$ broad, and 10\nobreakdashes--100\,Mpc deep. The ``forked tongue'' sky localization features due to the degeneracy of the sign of the binary inclination angle \citep{FirstTwoYears} are evident as narrow crevices running along the outside edges of the petals. The shape irresistibly suggests a tree ear fungus or a seed of the jacaranda tree.

The \acsu{O2} configuration \citep{LIGOObservingScenarios}, which may include \ac{LHO} and \ac{LLO} with improved sensitivity at a \ac{BNS} range of $\sim$100\,Mpc, as well as Advanced Virgo, leads to more compact and elaborate combinations of petal-shaped regions. In the most favorable three\nobreakdashes-detector cases where the area on the sky is localized to a single compact region, the reconstructed volume is a spindle a few degrees in radius and $\sim 100$\,Mpc long.

\section{Rapid volume reconstruction}

Although the reconstructed regions are highly structured, the posterior probability distribution \emph{along a given \acl{LOS}} is simple and generally unimodal; once again, a consequence of the Malmquist bias and the universal distribution of binary inclination angles.

This intuition leads us to suggest that the \emph{conditional distribution of distance} is well fit by an \emph{ansatz} whose location parameter $\hat\mu(\boldsymbol{n})$, scale $\hat\sigma(\boldsymbol{n})$, and normalization $\hat{N}(\boldsymbol{n})$ vary with sky location $\boldsymbol{n}$:
\begin{align}
    \label{eq:conditional-distance-distribution-ansatz}
    p(r | \boldsymbol{n}) &= \frac{\hat{N}(\boldsymbol{n})}{\sqrt{2\pi}\hat\sigma{(\boldsymbol{n})}} \exp\left[-\frac{\left(r - \hat\mu(\boldsymbol{n})\right)^2}{2\hat\sigma(\boldsymbol{n})^2}\right] r^2 \\
    \text{for } r &\geq 0. \nonumber
\end{align}
This form is equivalent to the product of a Gaussian likelihood and a uniform-in-volume prior. We show that this is a good fit in Section~6 of the \citetalias{GoingTheDistanceSupplement}.

The outputs of the \ac{LIGO}\nobreakdashes--Virgo localization pipelines are \acs{HEALPix} (\aclu{HEALPix}) all\nobreakdashes-sky images whose $N_\mathrm{pix}$ pixels give the posterior probability $\rho_i$ that the source is contained inside pixel $i$. We add three additional layers: $\hat\mu_i = \hat\mu(\boldsymbol{n}_i)$, $\hat\sigma_i = \hat\sigma(\boldsymbol{n}_i)$, and (for convenience) $\hat{N}_i = \hat{N}(\boldsymbol{n}_i)$. The first layer, $\rho_i$, is unchanged and still represents the \ac{2D} probability sky map.

The probability that a source is within pixel $i$ and at a distance between $r$ and $r + d r$ is $\rho_i$ times Eq.~(\ref{eq:conditional-distance-distribution-ansatz}):
\begin{equation}
    \label{eq:posterior-probability-distribution-ansatz}
    P(r, \boldsymbol{n}_i) \, d r = \rho_i \frac{\hat{N}_i}{\sqrt{2\pi}\hat\sigma_i} \exp\left[-\frac{\left(r - \hat\mu_i\right)^2}{2{\hat\sigma_i}^2}\right] r^2 \, d r.
\end{equation}
The sky map is normalized\footnote{There is no explicit area element because the pixels all have equal area.} such that
\begin{equation}
    \sum_{i\,=\,0}^{N-1} \rho_i = 1,
\end{equation}
so Eq.~(\ref{eq:posterior-probability-distribution-ansatz}) is also normalized such that
\begin{equation}
    \sum_{i\,=\,0}^{N-1} \int_0^\infty P(r, \boldsymbol{n}_i) \, d r = 1.
\end{equation}

The $r^2$ term is necessary in Eqs.~(\ref{eq:conditional-distance-distribution-ansatz},\,\ref{eq:posterior-probability-distribution-ansatz}) so that the probability density per unit volume vanishes at the origin. Eq.~(\ref{eq:posterior-probability-distribution-ansatz}) should be thought of as the probability distribution in spherical polar coordinates. If, however, one needs to perform a calculation in Cartesian coordinates, one converts using volume element, given by
\begin{equation}
    d V = r^2 \, d r \, \Delta\Omega = \frac{4\pi}{N_\mathrm{pix}} r^2 \, d r.
\end{equation}
The $r^2$ cancels in the resulting probability density per unit volume:
\begin{equation}
    \label{eq:probability-density-per-unit-volume}
    \frac{d P}{d V} = \rho_i \frac{N_\mathrm{pix}}{4\pi} \frac{\hat{N}_i}{\sqrt{2\pi}\hat\sigma_i} \exp\left[-\frac{\left(r - \hat\mu_i\right)^2}{2{\hat\sigma_i}^2}\right].
\end{equation}

Sky maps for compact binary merger candidates are produced by two codes with complementary sophistication and speed. The first is \acs{BAYESTAR}, which rapidly triangulates matched\nobreakdashes-filter estimates of the times, amplitudes, and phases on arrival at the \ac{GW} sites \citep{BAYESTAR}. The second is LALInference, which stochastically samples from sky location, distance, and component masses and spins \citep{LALInference}. Both methods directly sample the full \ac{3D} posterior probability distribution. The ansatz parameters are extracted using the method of moments as elaborated upon in Section~5 of the \citetalias{GoingTheDistanceSupplement}.

\section{Implications for early Advanced LIGO and Virgo}
\label{sec:implications}

\defcitealias{GCN18850}{LSC \& Virgo 2016}
We use this encoding to demonstrate the utility of the \ac{3D} structure of the \ac{GW} posteriors. \ac{LIGO} provided \ac{2D} localizations during \ac{O1} but did not calculate or distribute low-latency \ac{GW} distance estimates. (A directional distance estimate for GW151226 produced two weeks after the event (\citetalias{GCN18850}; \citealt{GW151226-PANSTARRS}) did help to rule out the redshift of a Pan-STARRS optical transient candidate.) Without a \ac{3D} sky map, one could have provided the horizon distance $r_\mathrm{H}$ calculated from the detectors' sensitivity or the effective distance $r_\mathrm{eff}$ based on the signal's \ac{SNR}. The new \ac{3D} sky maps allow us to find 90\% credible volumes that are 2--30 times smaller (10th to 90th percentile) than the volume within the 2D 90\% credible area and the horizon distance, or 1--7 times smaller than the volume within the 2D 90\% credible area and the effective distance.

However, the \ac{3D} localizations truly shine when we minimize not the volume, but rather the total exposure time required to observe every galaxy within the 90\% credible volume to a given flux limit. We neglect intrinsic scatter in absolute magnitude; the resulting conservative figure of merit allows us to focus on the effect of the distance posterior itself.

\begin{figure}
    \includegraphics[width=\columnwidth]{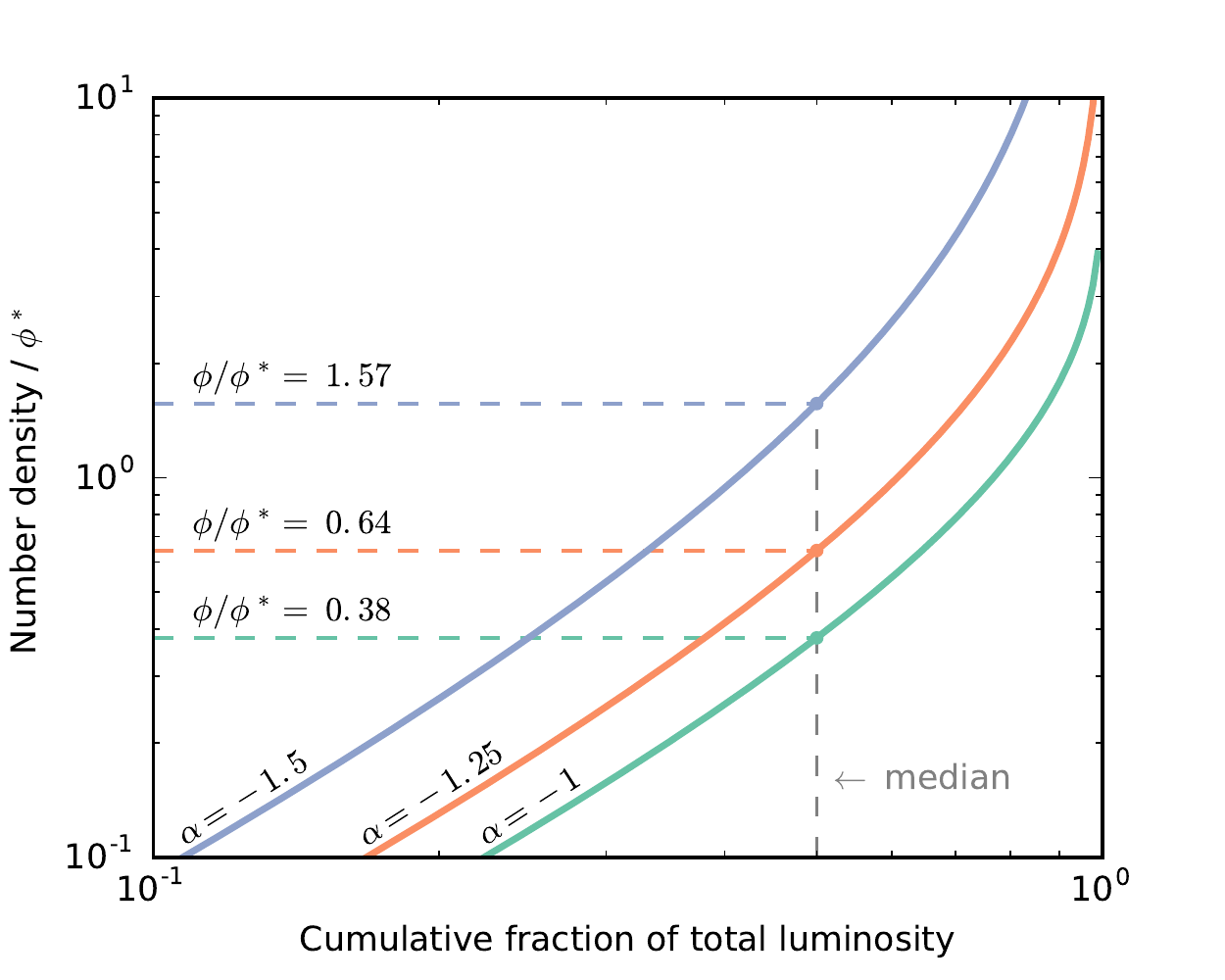}
    \caption{\label{fig:mass-quantiles}Tradeoff between completeness and the number of galaxies, assuming various Schechter luminosity functions. The horizontal axis is the cumulative fraction of the total luminosity in the local Universe. The vertical axis is the number density of galaxies in units of $\phi^*$.}
\end{figure}

If \ac{BNS} mergers have hosts that are similar to \acp{SGRB}, then their local rates are likely traced by a combination of recent star formation (measured by $B$-band luminosity) and stellar mass (measured by $K$-band luminosity; e.g. \citealt{2010ApJ...725.1202L,2013ApJ...769...56F}). As in \citet{GalaxyStrategy}, we attempt to mitigate these concerns as the limited completeness of galaxy catalogs by considering only the brightest galaxies. If we assume a $B$\nobreakdashes-band Schechter function with $\alpha=-1.25$, $\phi^*=1.2 \times 10^{-2}$\,$h^3$\,Mpc$^{-3}$, and $L^* = 1.2 \times 10^{10}$\,$h^{-2}$\,$L_\odot$ \citep{LongairGalaxyFormation} as a proxy for the \ac{BNS} merger rate, then we find that $2.8 \times 10^{-3}$~galaxies per Mpc$^3$ comprise half of the total luminosity (see Fig.~\ref{fig:mass-quantiles}). The areal density of galaxies out to a distance $r$ and with a luminosity greater than $L$ is
\begin{multline}
    N_\mathrm{gal} \approx 0.28 \text{ deg}^{-2} \\
    \left(\frac{r}{100\text{ Mpc}}\right)^3
    \left(\frac{\phi^*}{4 \times 10^{-3} \text{ Mpc}^{-3}}\right)
    \Gamma\left(\alpha + 1, \frac{L}{L^*}\right),
\end{multline}
at most a handful per deg$^2$ within distances that are relevant for \ac{BNS} mergers.

As a proof of concept, in the right panel of Fig.~\ref{fig:volume}, we show the potential host galaxies that are consistent with a simulated early Advanced \acs{LIGO} \ac{BNS} localization. We use the galaxy group catalog of \citet{2MASSGalaxyGroups}. This may not be an ideal galaxy catalog for \ac{GW} follow-up; among other reasons, being derived from 2MASS, its magnitudes trace mass rather than recent star formation. Better alternatives that should be available in the near future include the Census of the Local Universe (D.~Cook et~al. 2016, in preparation; \citealt{GalaxyStrategy}) and the Galaxy List for the Advanced Detector Era.\footnote{\url{http://aquarius.elte.hu/glade}} However, pending the availability of larger compilations of galaxy catalogs, it serves to illustrate our idea because it is $\sim∼$50\% complete out to $\sim 150$\,Mpc. Furthermore, incompleteness is encoded self-consistently by placing a variable bandwidth weighted kernel at the position of each galaxy.

Our control observing strategy takes all galaxies within a given 2D credible region, out to an optimal direction-independent distance inferred from the estimated mass of the source, the loudness of the signal, and the sensitivity of the detectors. It employs the same exposure time for every galaxy. This naive \ac{2+1D} construction is already far more sophisticated than the effective distance or horizon distance that could have been available in O1.

Different facilities demand different figures of merit. For \acp{FOV} $\ll 0.5 \text{ deg}^2$, most observations pick out single galaxies. In this case, the optimal selection of galaxies exploits both the upper and lower limits of the reconstructed volume, and the total exposure time is dependent on the number density of galaxies. For \acp{FOV} $\gg 0.5 \text{ deg}^2$, most observations pick out many galaxies; an exposure tuned to reach a fixed luminosity limit for a distant galaxy also captures all of the galaxies at intervening distances. In this case, we tune the exposure time to reach the most distant galaxy in each field, and the total exposure time is independent of the number density of galaxies. The instrument sensitivity is also important. In the source limited regime, the exposure time to reach a fixed limiting luminosity for a source at a distance $r$ varies as $r^2$. For sky background limited imaging, the time scales as $r^4$. The three cases of greatest interest are summarized below:
\begin{equation}
    \label{eq:fom}
    T = \min_{D: \int\limits_D P \, d V \,=\, 0.9} \begin{cases}
        \displaystyle\frac{T^*}{\omega} \displaystyle\int_D \left(\frac{r}{r^*}\right)^4 d\Omega & \substack{\displaystyle{\text{\emph{I}: large \acs{FOV},}} \\ \displaystyle{\text{sky limited}}} \\[2ex]
        T^* \phi \displaystyle\int_D \left(\frac{r}{r^*}\right)^2 d V & \substack{\displaystyle{\text{\emph{II}: small \acs{FOV},}} \\ \displaystyle{\text{source limited}}} \\[2ex]
        T^* \phi \displaystyle\int_D \left(\frac{r}{r^*}\right)^4 d V & \substack{\displaystyle{\text{\emph{III}: small \acs{FOV},}} \\ \displaystyle{\text{sky limited}.}}
    \end{cases}
\end{equation}
Here, $T^*$ is the time required to reach a fiducial flux limit for a source at a distance $r^*$, $\phi$ is the average number density of galaxies, and $\omega$ is the \ac{FOV} of the instrument in steradians. The integral is over solid angle (case~\emph{I}) or volume (cases~\emph{II} and~\emph{III}) and is minimized over all possible regions $D$ that contain 90\% posterior probability. Table~\ref{table:medians} presents the results of this exercise: volumes, numbers of galaxies, and exposure times. Each case typifies a search for a different \ac{EM} signature with a different kind of instrument:

\paragraph*{Case I} describes a \acl{KN} search with a synoptic optical survey instrument. We adopt the \ac{ZTF} our example and assume a \ac{FOV} of 47\,deg$^2$. Adopting $M_i = -13$ as the absolute magnitude of a \acl{KN}, we derive from Table~1 of \citet{KasliwalTwoDetectors} a fiducial \ac{ZTF} exposure time of $T^* \approx 6600$\,s for a limiting magnitude of $i \approx 23.5$ at a distance of $r^* = 200$\,Mpc. In \ac{O1}, a \acs{ZTF}\nobreakdashes-like survey would have taken $\sim$0.4\,hr to tile the entire region to an appropriate depth in exposures that average $\sim$2\,minutes in duration. The number of exposures decreases to a handful in the HLV configuration in \ac{O2}, but the exposure duration increases to 35\,minutes per field.

\paragraph*{Case II} models an afterglow search with the \emph{Swift} \ac{XRT} (considered in detail by \citealt{SwiftFollowupGWTransientsEvans}). Following \citet{KannerSwiftLIGOFollowUp}, we adopt a fiducial exposure time of $T^* = 100$\,s at $r^* = 200$\,Mpc. In \ac{O1}, we find an average exposure time of 7\,s per field, increasing to 25\,s per field in \ac{O2}. These are unrealistically short due to \emph{Swift}'s slew rate ($3\arcdeg$\,s$^{-1}$) and overhead (25\,s; \citealt{SwiftFollowupGWTransientsEvans}). However, even adding an overhead of $\sim$30\,s per galaxy, we suggest that a galaxy\nobreakdashes-targeted \emph{Swift} campaign could be accomplished within a few 90\,minute orbits.

\paragraph*{Case III} describes a \acl{KN} search with a traditional small \ac{FOV} optical telescope. Since we need to examine tens to hundreds of galaxies, we restrict ourselves to fully robotic telescopes dedicated to time\nobreakdashes-domain science and select the \ac{LCOGT} and Liverpool 2\,m telescopes as our models. Using the \ac{LCOGT} exposure time calculator,\footnote{\url{http://lcogt.net/files/etc/exposure_time_calculator.html}} we find an exposure time of $T^* \approx 2500$\,s to reach a depth of $i = 23.5$\,mag at $\text{\ac{SNR}}=5$ during half moon phase on the Spectral camera. In \ac{O1} we find an average exposure time of about 0.2\,minute, for a total exposure time of 0.4\,hr. Overhead would dominate. The average exposure time increases to $\sim$2.5\,minutes in \ac{O2}. The total exposure time increases dramatically from 0.4 to over 13.4\,hr due to the rapid increase in exposure time with distance and the modest increase in the number of galaxies. However, the total exposure time is still short enough that two 2\,m telescopes working in coordination could conceivably monitor all of the selected galaxies at a 1\nobreakdashes--2 night cadence.

\begin{deluxetable*}{rrrrr|rrr|rrr|rrr|rrr}
    \tablecolumns{17}
    \tablewidth{\textwidth}
    \tablecaption{\label{table:medians}Median Volumes and Exposure Times}
    \tablehead{
      \\[-2em]
        \colhead{} &
        \colhead{} &
        \colhead{} &
        \multicolumn{2}{c}{Range} &
        \multicolumn{3}{c}{} &
        \multicolumn{3}{c}{Large FOV/Sky} &
        \multicolumn{3}{c}{Small FOV/Source} &
        \multicolumn{3}{c}{Small FOV/Sky}
        \\
        \colhead{} &
        \colhead{} &
        \colhead{} &
        \multicolumn{2}{c}{(Mpc)} &
        \colhead{Vol. ($10^3$} &
        \colhead{No.} &
        \colhead{} &
        \multicolumn{3}{c}{Ex.: ZTF} &
        \multicolumn{3}{c}{Ex.: \emph{Swift} XRT} &
        \multicolumn{3}{c}{Ex.: LCOGT 2 m}
        \\
        \cline{4-5}
        \colhead{Run} &
        \colhead{Year} &
        \colhead{Net.\tablenotemark{a}} &
        \colhead{HL} &
        \colhead{V} &
        \colhead{Mpc$^3$)} &
        \colhead{Gal.} &
        \colhead{Red.\tablenotemark{b}} &
        \colhead{Tot.\tablenotemark{c}} &
        \colhead{Avg.\tablenotemark{d}} &
        \colhead{Red.\tablenotemark{b}} &
        \colhead{Tot.\tablenotemark{c}} &
        \colhead{Avg.\tablenotemark{d}} &
        \colhead{Red.\tablenotemark{b}} &
        \colhead{Tot.\tablenotemark{c}} &
        \colhead{Avg.\tablenotemark{d}} &
        \colhead{Red.\tablenotemark{b}}
    }
    \startdata
    \ac{O1} & 2015 & HL & 54 & --- &
    29 & 80 & 0.62 & 0.4 & 2 & 0.41 & 0.2 & 0.1 & 0.46 & 0.4 & 0.2 & 0.35 \\\tableline
    \multirow{2}{*}{\ac{O2}} & \multirow{2}{*}{2016} & HL & 108 & --- &
    324 & 906 & 0.59 & 6.9 & 23 & 0.38 & 6.9 & 0.4 & 0.42 & 56.2 & 2.6 & 0.31 \\
    & & HLV & 108 & 36 &
    56 & 156 & 0.75 & 1.5 & 35 & 0.52 & 1.5 & 0.4 & 0.58 & 13.4 & 2.5 & 0.43\\[-0.78em]
    \enddata
    \tablenotetext{a}{Network of \ac{GW} facilities that are in observing mode at the time of the event: LIGO Hanford~(H), LIGO Livingston~(L), or Virgo~(V).}
    \tablenotetext{b}{Reduction in optimal 3D strategy compared to naive \ac{2+1D} strategy}
    \tablenotetext{c}{Total exposure time in hours}
    \tablenotetext{d}{Average time per exposure in  minutes}
\end{deluxetable*}

\section{Discussion}
\label{sec:discussion}

To order to focus on the utility of distance and structure information, we neglected several details. \citet{LoudestGWEvents} address optimal selection of which \ac{GW} events to follow up. We set aside the feasibility of imaging tens to hundreds of targets in rapid succession with \emph{Swift}, though this is being implemented \citep{SwiftFollowupGWTransientsEvans}. We ignored the question of whether the \ac{XRT} exposure time can be finely tuned from one target to the next, though our results justify doing so if possible. We ignored variation in observability conditions such as Sun and Earth avoidance, weather, \acl{SAA} passages, and Moon phase, details which are better left to a facility\nobreakdashes-specific paper. We did not consider optimizing field selection given limited time resources, for which we refer the reader to \citet{ChanKilonovaDetectability}.

We did not address the significant issues of galaxy catalog completeness, construction of a galaxy prior from an incomplete galaxy catalog \citep{BayesianMultiMessenger}, or the suitability of any particular galaxy catalog, though targeting the brightest and most massive galaxies (those with $L \gtrsim L^*$) should simplify these concerns \citep{GalaxyStrategy}.

\citet{2MASSPhotoZLIGOOptimization} proposed using photometric redshifts to optimize \ac{GW} follow-up, improving completeness at the expense of accurate distances. Improved completeness mitigates the concern that \ac{BBH} mergers like GW150914, which are probably formed in low-metallicity environments \citep{GW150914-ASTRO}, may display host environment preferences similar to long \acp{GRB} and probably do not track with massive galaxies. However, our present work focuses on \ac{NS} binary mergers, which are expected to be found in fairly eclectic host environments due to the long time delay between formation and evolution of the binary and its \ac{GW}-driven inspiral into the \ac{LIGO} band. Therefore, the completeness of existing spectroscopic redshift surveys seems tolerable for our approach and for \acp{BNS}.

One possible concern for small-\ac{FOV} telescopes is the offsets that mergers may have from their host galaxies due to \ac{SN} kicks. \citet{LocationsShortGRBs} find that \acp{SGRB} have a median offsets of 4.5\,kpc from their hosts. For even an improbably nearby merger at $z = 0.005$ or a luminosity distance $r = 22$\,Mpc, the projected radius of the $24\arcmin$ \ac{XRT} \ac{FOV} is 74\,kpc, and of the $10\arcmin$ \ac{LCOGT} imager, 31\,kpc, encapsulating well over 90\% of \ac{SGRB} offsets.

In principal, our exposure time estimates should account not only for sky background, but additional image subtraction background due to the light of the host galaxy. Offsets should help here too because \acp{SGRB} are typically found at separations $>1.5$ times the effective radii of their host galaxies.

One particular advantage of galaxy-targeted searches is the reduction in false positives. In a magnitude-limited snapshot, \acp{SN} of Types~Ia, Ibc, and II are found in proportions of 68.6\%:4.3\%:27.1\% \citep{LickSNRateII}. Assuming a volumetric rate of $3\times10^{-5}$\,Mpc$^{-3}$\,yr$^{-1}$ \citep{LickSNRateII}, an average absolute magnitude of $-19$ over a duration of 1 week, and a limiting magnitude of 23.5, we calculate an areal rate of about 2 Type~Ia \acp{SN} per deg$^2$. A typical wide\nobreakdashes-field \ac{GW} follow\nobreakdashes-up campaign searching an area of $\sim 100$\,deg$^2$ will be contaminated by hundreds of \acp{SN}. However, if we consider only $10\arcmin \times 10\arcmin$ patches around 100 nearby galaxies, the on\nobreakdashes-sky footprint of $<$3\,deg$^2$ translates to a background of merely 6 \acp{SN}~Ia and 3 core-collapse \acp{SN}.

This points toward a new role in \ac{GW} follow\nobreakdashes-up for small-\ac{FOV}, large-aperture telescopes, in addition to and beyond their role of vetting candidates identified by synoptic surveys. Setting aside scheduling and proposal processes for the time being, suitable facilities for our strategy should have primary mirror diameters of 4--10\,m and optical or infrared imagers with $\sim10\arcmin$ \acp{FOV} that are either permanently installed at one of the foci or are rapidly deployable (e.g., mounted on Nasmyth platforms). Candidates include ALFOSC on the Nordic Optical Telescope, LMI on the Discovery Channel Telescope, WIRC on the Hale Telescope at Palomar, FourStar on Magellan, GMOS on Gemini North and South, FLAMINGOS-2 on Gemini South, FORS2 on VLT, LRIS at Keck, or GTC equipped with OSIRIS. As a pathfinder, we encourage deploying existing 2\,m class robotic optical telescopes in this manner during the early Advanced \acs{LIGO} and Virgo observing runs.

\providecommand{\acrolowercase}[1]{\lowercase{#1}}

\begin{acronym}
\acro{2D}[2D]{two\nobreakdashes-dimensional}
\acro{2+1D}[2+1D]{2+1\nobreakdashes--dimensional}
\acro{2MRS}[2MRS]{2MASS Redshift Survey}
\acro{3D}[3D]{three\nobreakdashes-dimensional}
\acro{2MASS}[2MASS]{Two Micron All Sky Survey}
\acro{AdVirgo}[AdVirgo]{Advanced Virgo}
\acro{AMI}[AMI]{Arcminute Microkelvin Imager}
\acro{AGN}[AGN]{active galactic nucleus}
\acroplural{AGN}[AGN\acrolowercase{s}]{active galactic nuclei}
\acro{aLIGO}[aLIGO]{Advanced \acs{LIGO}}
\acro{ASKAP}[ASKAP]{Australian \acl{SKA} Pathfinder}
\acro{ATCA}[ATCA]{Australia Telescope Compact Array}
\acro{ATLAS}[ATLAS]{Asteroid Terrestrial-impact Last Alert System}
\acro{BAT}[BAT]{Burst Alert Telescope\acroextra{ (instrument on \emph{Swift})}}
\acro{BATSE}[BATSE]{Burst and Transient Source Experiment\acroextra{ (instrument on \acs{CGRO})}}
\acro{BAYESTAR}[BAYESTAR]{BAYESian TriAngulation and Rapid localization}
\acro{BBH}[BBH]{binary black hole}
\acro{BHBH}[BHBH]{\acl{BH}\nobreakdashes--\acl{BH}}
\acro{BH}[BH]{black hole}
\acro{BNS}[BNS]{binary neutron star}
\acro{CARMA}[CARMA]{Combined Array for Research in Millimeter\nobreakdashes-wave Astronomy}
\acro{CASA}[CASA]{Common Astronomy Software Applications}
\acro{CFH12k}[CFH12k]{Canada--France--Hawaii $12\,288 \times 8\,192$ pixel CCD mosaic\acroextra{ (instrument formerly on the Canada--France--Hawaii Telescope, now on the \ac{P48})}}
\acro{CRTS}[CRTS]{Catalina Real-time Transient Survey}
\acro{CTIO}[CTIO]{Cerro Tololo Inter-American Observatory}
\acro{CBC}[CBC]{compact binary coalescence}
\acro{CCD}[CCD]{charge coupled device}
\acro{CDF}[CDF]{cumulative distribution function}
\acro{CGRO}[CGRO]{Compton Gamma Ray Observatory}
\acro{CMB}[CMB]{cosmic microwave background}
\acro{CRLB}[CRLB]{Cram\'{e}r\nobreakdashes--Rao lower bound}
\acro{cWB}[\acrolowercase{c}WB]{Coherent WaveBurst}
\acro{DASWG}[DASWG]{Data Analysis Software Working Group}
\acro{DBSP}[DBSP]{Double Spectrograph\acroextra{ (instrument on \acs{P200})}}
\acro{DCT}[DCT]{Discovery Channel Telescope}
\acro{DECAM}[DECam]{Dark Energy Camera\acroextra{ (instrument on the Blanco 4\nobreakdashes-m telescope at \acs{CTIO})}}
\acro{DES}[DES]{Dark Energy Survey}
\acro{DFT}[DFT]{discrete Fourier transform}
\acro{EM}[EM]{electromagnetic}
\acro{ER8}[ER8]{eighth engineering run}
\acro{FD}[FD]{frequency domain}
\acro{FAR}[FAR]{false alarm rate}
\acro{FFT}[FFT]{fast Fourier transform}
\acro{FIR}[FIR]{finite impulse response}
\acro{FITS}[FITS]{Flexible Image Transport System}
\acro{FLOPS}[FLOPS]{floating point operations per second}
\acro{FOV}[FOV]{field of view}
\acroplural{FOV}[FOV\acrolowercase{s}]{fields of view}
\acro{FTN}[FTN]{Faulkes Telescope North}
\acro{FWHM}[FWHM]{full width at half-maximum}
\acro{GBM}[GBM]{Gamma-ray Burst Monitor\acroextra{ (instrument on \emph{Fermi})}}
\acro{GCN}[GCN]{Gamma-ray Coordinates Network}
\acro{GMOS}[GMOS]{Gemini Multi-object Spectrograph\acroextra{ (instrument on the Gemini telescopes)}}
\acro{GRB}[GRB]{gamma-ray burst}
\acro{GSC}[GSC]{Gas Slit Camera}
\acro{GSL}[GSL]{GNU Scientific Library}
\acro{GTC}[GTC]{Gran Telescopio Canarias}
\acro{GW}[GW]{gravitational wave}
\acro{HAWC}[HAWC]{High\nobreakdashes-Altitude Water \v{C}erenkov Gamma\nobreakdashes-Ray Observatory}
\acro{HCT}[HCT]{Himalayan Chandra Telescope}
\acro{HEALPix}[HEALP\acrolowercase{ix}]{Hierarchical Equal Area isoLatitude Pixelization}
\acro{HEASARC}[HEASARC]{High Energy Astrophysics Science Archive Research Center}
\acro{HETE}[HETE]{High Energy Transient Explorer}
\acro{HFOSC}[HFOSC]{Himalaya Faint Object Spectrograph and Camera\acroextra{ (instrument on \acs{HCT})}}
\acro{HMXB}[HMXB]{high\nobreakdashes-mass X\nobreakdashes-ray binary}
\acroplural{HMXB}[HMXB\acrolowercase{s}]{high\nobreakdashes-mass X\nobreakdashes-ray binaries}
\acro{HSC}[HSC]{Hyper Suprime\nobreakdashes-Cam\acroextra{ (instrument on the 8.2\nobreakdashes-m Subaru telescope)}}
\acro{IACT}[IACT]{imaging atmospheric \v{C}erenkov telescope}
\acro{IIR}[IIR]{infinite impulse response}
\acro{IMACS}[IMACS]{Inamori-Magellan Areal Camera \& Spectrograph\acroextra{ (instrument on the Magellan Baade telescope)}}
\acro{IMR}[IMR]{inspiral-merger-ringdown}
\acro{IPAC}[IPAC]{Infrared Processing and Analysis Center}
\acro{IPN}[IPN]{InterPlanetary Network}
\acro{IPTF}[\acrolowercase{i}PTF]{intermediate \acl{PTF}}
\acro{ISM}[ISM]{interstellar medium}
\acro{ISS}[ISS]{International Space Station}
\acro{KAGRA}[KAGRA]{KAmioka GRAvitational\nobreakdashes-wave observatory}
\acro{KDE}[KDE]{kernel density estimator}
\acro{KN}[KN]{kilonova}
\acroplural{KN}[KNe]{kilonovae}
\acro{LAT}[LAT]{Large Area Telescope}
\acro{LCOGT}[LCOGT]{Las Cumbres Observatory Global Telescope}
\acro{LHO}[LHO]{\ac{LIGO} Hanford Observatory}
\acro{LIB}[LIB]{LALInference Burst}
\acro{LIGO}[LIGO]{Laser Interferometer \acs{GW} Observatory}
\acro{llGRB}[\acrolowercase{ll}GRB]{low\nobreakdashes-luminosity \ac{GRB}}
\acro{LLOID}[LLOID]{Low Latency Online Inspiral Detection}
\acro{LLO}[LLO]{\ac{LIGO} Livingston Observatory}
\acro{LMI}[LMI]{Large Monolithic Imager\acroextra{ (instrument on \ac{DCT})}}
\acro{LOFAR}[LOFAR]{Low Frequency Array}
\acro{LOS}[LOS]{line of sight}
\acroplural{LOS}[LOSs]{lines of sight}
\acro{LMC}[LMC]{Large Magellanic Cloud}
\acro{LSB}[LSB]{long, soft burst}
\acro{LSC}[LSC]{\acs{LIGO} Scientific Collaboration}
\acro{LSO}[LSO]{last stable orbit}
\acro{LSST}[LSST]{Large Synoptic Survey Telescope}
\acro{LT}[LT]{Liverpool Telescope}
\acro{LTI}[LTI]{linear time invariant}
\acro{MAP}[MAP]{maximum a posteriori}
\acro{MBTA}[MBTA]{Multi-Band Template Analysis}
\acro{MCMC}[MCMC]{Markov chain Monte Carlo}
\acro{MLE}[MLE]{\ac{ML} estimator}
\acro{ML}[ML]{maximum likelihood}
\acro{MOU}[MOU]{memorandum of understanding}
\acroplural{MOU}[MOUs]{memoranda of understanding}
\acro{MWA}[MWA]{Murchison Widefield Array}
\acro{NED}[NED]{NASA/IPAC Extragalactic Database}
\acro{NSBH}[NSBH]{neutron star\nobreakdashes--black hole}
\acro{NSBH}[NSBH]{\acl{NS}\nobreakdashes--\acl{BH}}
\acro{NSF}[NSF]{National Science Foundation}
\acro{NSNS}[NSNS]{\acl{NS}\nobreakdashes--\acl{NS}}
\acro{NS}[NS]{neutron star}
\acro{O1}[O1]{\acl{aLIGO}'s first observing run}
\acro{O2}[O2]{\acl{aLIGO}'s second observing run}
\acro{oLIB}[\acrolowercase{o}LIB]{Omicron+\acl{LIB}}
\acro{OT}[OT]{optical transient}
\acro{P48}[P48]{Palomar 48~inch Oschin telescope}
\acro{P60}[P60]{robotic Palomar 60~inch telescope}
\acro{P200}[P200]{Palomar 200~inch Hale telescope}
\acro{PC}[PC]{photon counting}
\acro{PESSTO}[PESSTO]{Public ESO Spectroscopic Survey of Transient Objects}
\acro{PSD}[PSD]{power spectral density}
\acro{PSF}[PSF]{point-spread function}
\acro{PS1}[PS1]{Pan\nobreakdashes-STARRS~1}
\acro{PTF}[PTF]{Palomar Transient Factory}
\acro{QUEST}[QUEST]{Quasar Equatorial Survey Team}
\acro{RAPTOR}[RAPTOR]{Rapid Telescopes for Optical Response}
\acro{REU}[REU]{Research Experiences for Undergraduates}
\acro{RMS}[RMS]{root mean square}
\acro{ROTSE}[ROTSE]{Robotic Optical Transient Search}
\acro{S5}[S5]{\ac{LIGO}'s fifth science run}
\acro{S6}[S6]{\ac{LIGO}'s sixth science run}
\acro{SAA}[SAA]{South Atlantic Anomaly}
\acro{SHB}[SHB]{short, hard burst}
\acro{SHGRB}[SHGRB]{short, hard \acl{GRB}}
\acro{SKA}[SKA]{Square Kilometer Array}
\acro{SMT}[SMT]{Slewing Mirror Telescope\acroextra{ (instrument on \acs{UFFO} Pathfinder)}}
\acro{SNR}[S/N]{signal\nobreakdashes-to\nobreakdashes-noise ratio}
\acro{SSC}[SSC]{synchrotron self\nobreakdashes-Compton}
\acro{SDSS}[SDSS]{Sloan Digital Sky Survey}
\acro{SED}[SED]{spectral energy distribution}
\acro{SGRB}[SGRB]{short \acl{GRB}}
\acro{SN}[SN]{supernova}
\acroplural{SN}[SN\acrolowercase{e}]{supernova}
\acro{SNIa}[\acs{SN}\,I\acrolowercase{a}]{Type~Ia \ac{SN}}
\acroplural{SNIa}[\acsp{SN}\,I\acrolowercase{a}]{Type~Ic \acp{SN}}
\acro{SNIcBL}[\acs{SN}\,I\acrolowercase{c}\nobreakdashes-BL]{broad\nobreakdashes-line Type~Ic \ac{SN}}
\acroplural{SNIcBL}[\acsp{SN}\,I\acrolowercase{c}\nobreakdashes-BL]{broad\nobreakdashes-line Type~Ic \acp{SN}}
\acro{SVD}[SVD]{singular value decomposition}
\acro{TAROT}[TAROT]{T\'{e}lescopes \`{a} Action Rapide pour les Objets Transitoires}
\acro{TDOA}[TDOA]{time delay on arrival}
\acroplural{TDOA}[TDOA\acrolowercase{s}]{time delays on arrival}
\acro{TD}[TD]{time domain}
\acro{TOA}[TOA]{time of arrival}
\acroplural{TOA}[TOA\acrolowercase{s}]{times of arrival}
\acro{TOO}[TOO]{target\nobreakdashes-of\nobreakdashes-opportunity}
\acroplural{TOO}[TOO\acrolowercase{s}]{targets of opportunity}
\acro{UFFO}[UFFO]{Ultra Fast Flash Observatory}
\acro{UHE}[UHE]{ultra high energy}
\acro{UVOT}[UVOT]{UV/Optical Telescope\acroextra{ (instrument on \emph{Swift})}}
\acro{VHE}[VHE]{very high energy}
\acro{VISTA}[VISTA@ESO]{Visible and Infrared Survey Telescope}
\acro{VLA}[VLA]{Karl G. Jansky Very Large Array}
\acro{VLT}[VLT]{Very Large Telescope}
\acro{VST}[VST@ESO]{\acs{VLT} Survey Telescope}
\acro{WAM}[WAM]{Wide\nobreakdashes-band All\nobreakdashes-sky Monitor\acroextra{ (instrument on \emph{Suzaku})}}
\acro{WCS}[WCS]{World Coordinate System}
\acro{WSS}[w.s.s.]{wide\nobreakdashes-sense stationary}
\acro{XRF}[XRF]{X\nobreakdashes-ray flash}
\acroplural{XRF}[XRF\acrolowercase{s}]{X\nobreakdashes-ray flashes}
\acro{XRT}[XRT]{X\nobreakdashes-ray Telescope\acroextra{ (instrument on \emph{Swift})}}
\acro{ZTF}[ZTF]{Zwicky Transient Facility}
\end{acronym}
 
\acknowledgements
We thank the Aspen Center for Physics and NSF grant \#1066293 for hospitality during the conception, writing, and editing of this paper. We thank P.~Shawhan and F.~Tombesi for detailed feedback on the manuscript. Supplementary material, including a sample of reconstructed \ac{GW} volume \acs{FITS} files, will be made available at \url{https://dcc.ligo.org/P1500071/public/html}. See the \citetalias{GoingTheDistanceSupplement} \citep{GoingTheDistanceSupplement} in the journal for more details. This is \acs{LIGO} document P1500071\nobreakdashes-v7.

\bibliographystyle{aasjournal}

\end{document}